# Can Plants Grow on the Moon and Mars: Seed Germination Enhancement Using Magnesium Oxide-Coated Halloysite Nanotubes (MgO-HNTs)

Zeinab Jabbari Velisdeh [1], David K. Mills [2*]

[1]Molecular Science and Nanotechnology, Louisiana Tech University, Ruston, LA. 71272
[2]OrganicNANO, Monroe, LA. 71201-3045

[*]Corresponding author
David K. Mills, Ph.D.
OrganicNANO
Monroe, LA 71201-3045
E-mail: david@minerva319.com

**Abstract**

This study examines the application of metal-coated nanotubes, specifically magnesium oxide–coated halloysite nanotubes (MgO-HNTs), to enhance seed germination and early plant development under Earth, lunar, and Martian soil conditions. MgO-HNTs were synthesized through an electrodeposition process and characterized by scanning electron microscopy (SEM) to confirm successful surface modification. Growth experiments using Heirloom Cherry Tomato and Golden Tomato seeds were conducted under hydroponic and soil-based conditions and subsequently extended to lunar and Martian regolith simulants. A Response Surface Methodology (RSM) approach, based on a Box–Behnken Design, was used to evaluate the effects of temperature, MgO-HNT concentration, and light duration on multiple growth responses. Seedling length and the root length stress tolerance index (RLSI) were the most responsive indicators of MgO-HNT treatment. Optimal conditions (25 °C, 12 h light exposure, 100 mg/L MgO-HNTs) produced the greatest increases in root and shoot length in Earth soil simulants. Validation experiments in extraterrestrial regolith analogs showed that MgO-HNTs supported germination, root penetration, and overall seedling vigor under nutrient-limited and high-stress conditions. Lunar regolith exhibited maximal root development at 100 mg/L, whereas Martian regolith showed optimal growth at 10 mg/L, reflecting differences in mineral composition and oxidative characteristics. These findings indicate that MgO-HNTs can enhance early plant development across both terrestrial and extraterrestrial substrates. The results support the potential use of MgO-HNT–based amendments within in-situ resource utilization (ISRU) frameworks to advance sustainable plant cultivation for future agricultural systems on Earth and in space.

1. Introduction

Global agricultural systems face increasing pressure to support the demands of a growing population while simultaneously addressing climate change, soil degradation, and resource scarcity [1,2]. Conventional practices reliant on synthetic fertilizers, monoculture, and intensive land use have contributed to nutrient depletion, loss of arable land, and environmental contamination [3]. These challenges have accelerated the transition toward sustainable agriculture, which emphasizes long-term productivity, ecological stability, and efficient resource use [4]. Recent developments in sustainable agriculture increasingly incorporate emerging technologies—such as nanomaterials and precision farming—to improve nutrient delivery, soil structure, and plant tolerance to abiotic stress [5-7]. At the same time, growing interest in extraterrestrial colonization has highlighted the need for closed-loop agricultural systems capable of supporting plant growth under extreme conditions, including nutrient-poor substrates and limited water availability [8,9]. Nanomaterials, defined as materials with at least one dimension between 1 and 100 nm, exhibit physicochemical properties such as high surface reactivity and enhanced stability that are not observed in bulk materials [10-12]. In agriculture, these features enable improved nutrient-use efficiency, soil conditioning, and enhanced plant resilience. Engineered nanomaterials have been shown to promote seed germination, stimulate early root and shoot development, and increase nutrient uptake. For example, zinc oxide and titanium dioxide nanoparticles can increase metabolic activity and support early growth in several plant species [13-14]. Nanomaterials are also used as nanofertilizers and nanocarriers, providing controlled nutrient release and reducing losses associated with conventional fertilizers [15]. Additional studies report increased tolerance to abiotic stresses—including drought and nutrient deficiency—through mechanisms related to enhanced antioxidant activity and nutrient availability [16-17]. These benefits extend beyond terrestrial systems, with nanomaterials proposed for plant cultivation in controlled-environment and extraterrestrial agriculture where efficient nutrient delivery and stress mitigation are critical [8-9].

Halloysite nanotubes (HNTs) are naturally occurring aluminosilicate nanomaterials composed of alternating $SiO_2$ and $Al_2O_3$ layers rolled into hollow tubular structures [18]. Their widespread natural availability in several countries makes them a low-cost and accessible alternative to synthetic nanotubes [18]. Typical HNTs measure 500–1500 nm in length, with external diameters of 40–70 nm and lumen diameters of 10–30 nm [19]. This tubular morphology provides high surface area and internal cavity volume suitable for loading and controlled release of active compounds. A defining feature of HNTs is their dual surface chemistry: negatively charged siloxane groups on the outer surface and positively charged aluminol groups within the lumen (Figure 1) [20]. This bipolar structure enables selective functionalization and dual loading of oppositely charged molecules. The general chemical composition of HNTs is summarized in Table 1 [21]. HNTs offer several advantages for agricultural applications, including biocompatibility, low toxicity, environmental safety, and substantially lower cost than carbon nanotubes, which often require additional modification to reduce cytotoxicity [22-25]. Their chemical stability and ability to form stable aqueous dispersions further support their use in soil amendments, nanofertilizers, and micronutrient carriers [26]. Based on these properties, HNTs were selected as the nanocarrier for this study.

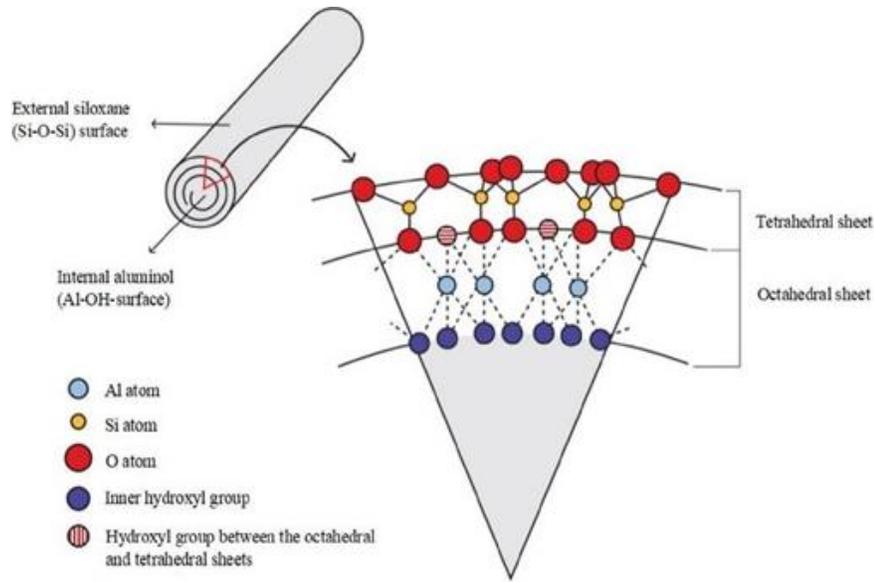

**Figure 1**: Schematic Structure of a Halloysite Nanotube (HNT), Showing the Aluminol (Al–OH) Groups Lining the Inner Lumen and Siloxane (Si–O–Si) Groups on the Outer Surface [20].

**Table 1:** Chemical composition of halloysite nanotubes (HNTs) [21].

| Composition | % (wt.) |
|---|---|
| $SiO_2$ | 46.15 |
| $Al_2O_3$ | 38.7 |
| $TiO_2$ | 0.004 |
| $Fe_2O_3$ | 0.05 |
| $CaO$ | 0.192 |
| $MgO$ | 0.035 |
| $K_2O$ | 0.03 |
| $Na_2O$ | 0.04 |
| $P_2O_5$ | 0.044 |
| $MnO$ | 0.001 |
| $H_2O$ | 14.6 |

Metal and metal oxide coatings enhance the functional performance of halloysite nanotubes (HNTs), improving their chemical reactivity, antimicrobial activity, nutrient delivery capacity, and environmental stability [27-29]. Coatings with magnesium, iron, zinc, copper, or silver have been investigated for agricultural applications, enabling HNTs to act as carriers for essential micronutrients and stress-mitigating agents. Magnesium-coated HNTs are of particular interest because magnesium is required for chlorophyll synthesis and enzymatic function [28]. Zinc- and iron-coated HNTs have also been reported to improve germination, nutrient uptake, and root development [30], while silver and copper coatings provide antimicrobial effects that may help reduce soilborne pathogens [31]. Several coating techniques are available, including sol–gel processing, wet impregnation, chemical precipitation, thermal treatment, and electrodeposition [32]. Electrodeposition was selected in this study to produce magnesium oxide–coated HNTs (MgO-HNTs) due to its controllability and compatibility with aqueous, environmentally benign processing.

Magnesium is an essential macronutrient involved in photosynthesis, enzyme activation, energy metabolism, and nutrient transport. As the central atom of the chlorophyll molecule, it enables efficient light absorption and energy conversion [33]. Magnesium deficiency, common in acidic or sandy soils and in intensively fertilized systems, leads to chlorosis, reduced root growth, and lower crop productivity [34]. Elemental magnesium is highly reactive and therefore unsuitable for direct agricultural use due to rapid oxidation and instability under ambient conditions [35]. Magnesium oxide (MgO), a stable and water-insoluble compound, is consequently preferred as a practical source of magnesium. MgO is valued for its thermal stability, low toxicity, and basic pH, which can help neutralize acidic soils and improve nutrient availability [36]. At the nanoscale, MgO exhibits increased surface area and reactivity, improving dispersion, interaction with soil particles, and nutrient delivery efficiency. Nano-MgO has been reported to enhance seed germination, root elongation, biomass accumulation, and plant tolerance to nutrient deficiency and abiotic stress [37,38]. MgO nanoparticles also demonstrate mild antimicrobial activity, which may contribute to improved rhizosphere conditions [37]. A key agronomic function of MgO is its ability to buffer soil acidity: acting as a liming agent, it reduces hydrogen ion concentration in the root zone and increases the availability of phosphorus, calcium, and other nutrients [39]. Its combined role as a magnesium source and soil conditioner supports its integration into precision agriculture platforms. The MgO coating mechanism exploits the surface chemistry of HNTs. Negatively charged siloxane (Si–O–Si) groups on the outer wall promote electrostatic attachment of MgO particles or precursor ions, whereas positively charged aluminol (Al–OH) groups inside the lumen may support partial internal loading. This configuration enables both rapid and sustained nutrient release. A schematic representation of magnesium deposition onto HNT surfaces is shown in Figure 2. MgO-functionalized nanocarriers—such as silica-, biochar-, and carbon-based systems—have been investigated for improving nutrient retention and use efficiency in degraded soils [40,41]. The use of naturally occurring HNTs offers a cost-effective and environmentally compatible alternative, requiring minimal processing and supporting uniform, stable coatings due to their intrinsic surface charge and tubular structure. The resulting MgO-HNT system is particularly relevant for low-input or resource-limited agricultural environments, where fertilizer loss, soil acidity, and moisture limitations restrict crop productivity. The multifunctionality of MgO-HNTs also aligns with the needs of extraterrestrial agriculture. In lunar or Martian regolith, where nutrient content, water availability, and soil structure present substantial constraints, controlled nutrient release and moisture retention are critical. MgO-HNTs are compatible with in-situ resource utilization (ISRU) strategies due to their low mass, stability, and adaptability, making them strong candidates for plant cultivation in closed-loop systems.

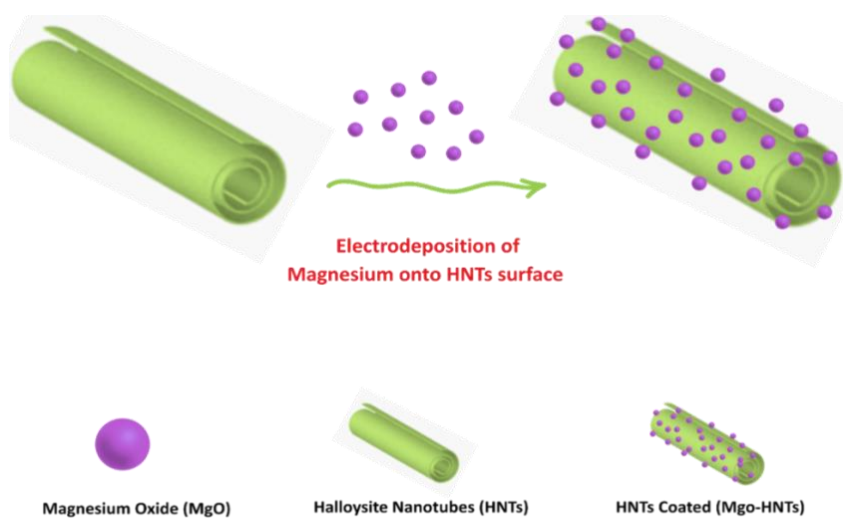

**Figure 2**: Schematic Illustration of Magnesium Electrodeposition onto Halloysite Nanotube (HNT) Surfaces. (This figure was created by the author using Canva Premium. Usage and reproduction are permitted for academic, research, and publication purposes under Canva's content license.)

Seed germination and early root development are critical phases that determine crop establishment, resilience, and yield potential. These stages require adequate water, oxygen, nutrients, and favorable physicochemical conditions; when any of these factors are limited, germination is delayed and root growth restricted, reducing overall plant performance [42]. Nutrient deficiency is a major constraint during germination, particularly in degraded, sandy, or acidic soils. Magnesium deficiency, in particular, impairs chlorophyll formation and enzymatic activity during early development [43]. Abiotic stresses—including drought, salinity, temperature extremes, and low pH—further hinder root elongation and cellular activity essential for seedling establishment [44]. Conventional fertilization often provides limited benefit during early growth because nutrients may not be available at the precise time or location needed by emerging seedlings. Losses through leaching or volatilization, especially in porous or poorly structured soils, further reduce fertilizer efficiency [45].

These limitations highlight the need for targeted nutrient delivery systems capable of supporting early root–soil interactions. Nanotechnology-based approaches have shown promise for improving germination and root development by enhancing water retention, moderating soil pH, and supplying nutrients directly to the root zone [46]. Nanomaterials interact closely with seed surfaces and root tips, promoting cellular activity and increasing tolerance to environmental stressors. In this study, early-stage plant performance was evaluated using several physiological indicators, including seedling length, germination percentage, root-to-shoot ratio, seedling vigor index (SVI), and stress tolerance indices. These metrics provided a comprehensive assessment of germination success and seedling resilience under nutrient-limited conditions. Response Surface Methodology (RSM) was employed to evaluate the combined effects of multiple experimental variables and to identify optimal conditions for early plant development. Unlike one-factor-at-a-time approaches, RSM accounts for interactions among variables and efficiently models nonlinear response behavior [47]. A Box–Behnken Design (BBD) was selected due to its suitability for three-factor optimization and its reduced number of required experimental runs. The design generated a quadratic polynomial model capable of estimating main effects, interaction effects, and second-order terms [48]. Model adequacy was assessed using analysis of variance (ANOVA), lack-of-fit tests, and coefficient of determination ($R^2$). Response surfaces and contour plots were generated to visualize factor interactions and determine optimal conditions.

This study evaluates magnesium oxide–coated halloysite nanotubes (MgO-HNTs) as a strategy to enhance plant establishment in nutrient-limited terrestrial and extraterrestrial substrates, contributing to broader efforts to advance sustainable agriculture on Earth and in space.

## 2. Materials and Methods

### 2.1. Chemicals and Reagents

Magnesium oxide (MgO, CAS: 1309-48-4), zinc powder (Zn), and halloysite nanoclay (LOT: 685445-500G, CAS: 1332-58-7) were all obtained from Sigma-Aldrich, St. Louis, MO, USA. Ethanol (≥99.5%, CAS: 64-17-5) was used for surface cleaning and sample preparation steps. Deionized (DI) water was used for all experimental procedures and was produced in the laboratory using deionization system. For plant growth experiments, Heirloom Cherry Tomato and Golden Tomato seeds were obtained from the AeroGarden Pod Kit, purchased via Amazon. Growth substrates included sterilized Earth soil and two types of extraterrestrial soil simulants. Martian regolith simulant (MMS-1 Fine Grade and MMS-1 Unsorted Grade) was sourced from The Martian Garden (Austin, TX, USA) and designed to replicate the mineral composition and texture of the Martian surface. Lunar regolith simulant (LHS-1D Dust Lunar Highlands) was obtained from Exolith Lab (University of Central Florida, Orlando, FL, USA) and formulated to simulate the lunar highlands' dusty soil environment. These substrates were used to mimic nutrient-deficient and harsh extraterrestrial conditions for seed germination and early plant development studies.

### 2.2. Synthesis of MgO-HNTs

Metal-coated halloysite nanotubes were synthesized using an electrodeposition method to functionalize HNTs with magnesium oxide (MgO), as shown in Figure 3. The coating process was performed in a 1000 mL glass electrolysis vessel containing 700 mL of deionized water, maintained at 85 °C. A total of 350 mg of halloysite nanoclay was dispersed in the heated solution, followed by the addition of 141.07 mg of magnesium oxide (for MgO-HNTs). Continuous stirring was maintained using a magnetic stir bar to ensure homogeneity. The electrolysis setup included two platinum-coated titanium mesh electrodes placed approximately 2 inches apart. Prior to use, the electrodes were polished with silicon carbide abrasive paper and ultrasonicated in distilled water for 10 minutes to remove surface contaminants. The electrodes were connected to a DC power supply operated at 20 V. To prevent localized buildup of materials and promote uniform deposition, the polarity of the electrodes was reversed every 5 minutes over a 30-minute deposition period (six cycles total). After electrodeposition, the suspension was allowed to rest for 5 minutes, after which the supernatant was decanted. Each coated formulation was then washed three times with deionized water and centrifuged at 2000 rpm for 5 minutes using an Eppendorf Centrifuge 5702 R (Germany) to separate unreacted particles. The final pellet was dried at 37 °C and stored in a desiccator for further experimental use.

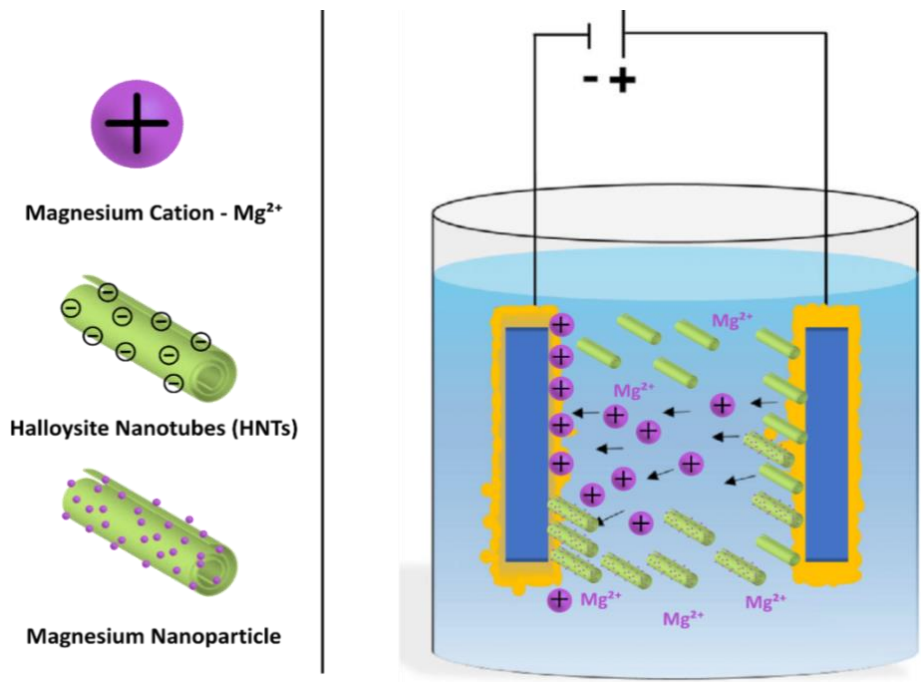

**Figure 3**: Schematic diagram of the electrodeposition process used to coat HNTs with metal oxides (MgO - Mg2+). (Created by the author using Canva Premium; reproduction permitted under Canva's content license.)

### 2.3. Material Characterization

The synthesized magnesium oxide–coated halloysite nanotubes (MgO-HNTs) were characterized to verify successful coating and assess structural and elemental properties. Surface morphology was examined using a Hitachi S-4800 Field Emission Scanning Electron Microscope (FE-SEM), and elemental composition was analyzed using an EDAX Energy Dispersive X-ray Spectroscopy (EDS) system integrated with the FE-SEM. SEM provided microstructural imaging of the coated nanotubes, while EDS confirmed magnesium incorporation on the HNT surfaces.

### 2.4. Soil and Regolith Preparation

This study involved a two-phase soil preparation and testing approach to evaluate the performance of MgO-HNTs in enhancing seed germination and root development. Commercially available agricultural soil was used as the baseline substrate for optimizing plant growth conditions. The soil air-dried, passed through a 2 mm mesh sieve to remove large particles and debris, and stored in clean containers at room temperature. This low-nutrient soil served as a control medium to test the effects of MgO-HNTs concentration, light intensity, and temperature on early-stage plant development. Following optimization in Earth soil, the best-performing conditions were applied to lunar and Martian regolith simulants to evaluate plant performance in extraterrestrial environments. Regolith simulants were used as received to preserve their mineral integrity. Before use, simulants were manually homogenized and air-dried under ambient laboratory conditions.

### 2.5. Seed Germination and Growth Conditions

Seed germination and plant growth experiments were conducted using controlled hydroponic and soil-based systems to evaluate the effects of MgO-coated halloysite nanotubes (MgO-HNTs) on early-stage tomato development. Two tomato varieties—Heirloom Cherry Tomato and Golden Tomato—were used throughout the study. Seeds were first germinated in a 12-pod hydroponic system equipped with built-in LED grow lights (Figure 4A), which provided consistent illumination and hydration during early sprouting under sterile conditions. Seedlings were monitored daily to ensure uniform development. After germination, seedlings were transferred to a controlled growth chamber (Figure 4B), where environmental conditions were maintained at 30 °C with an 8-hour photoperiod and ambient laboratory humidity (~70% RH).

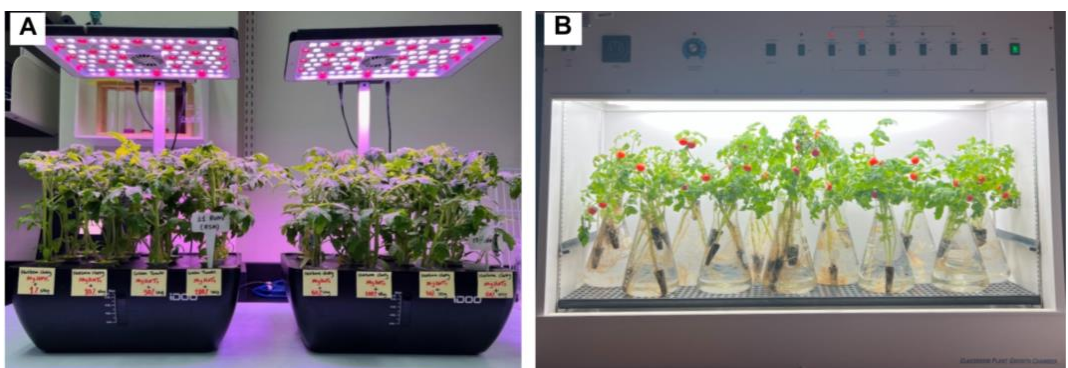

**Figure 4.** Seed germination and growth systems used in this study. (A) Hydroponic 12-pod system with integrated LED lighting for initial seed sprouting. (B) Controlled-environment growth chamber used for subsequent seedling development under regulated temperature, humidity, and light conditions.

### 2.6. Treatment Methods

To evaluate the effects of MgO-HNTs on early plant development, two application methods—foliar spraying and soil injection—were employed. For foliar application, aqueous suspensions of MgO-HNTs were prepared in deionized water at concentrations of 0%, 1%, 3%, 5%, and 10% (w/v) and applied directly to the leaves using sterile spray bottles. Treatments were administered twice daily at 12-hour intervals. For soil injection, MgO-HNT suspensions were prepared at concentrations of 0.1, 1, 10, 50, and 100 mg per plant and delivered to the root zone using sterile syringes; parallel treatments using pure MgO at identical concentrations were included for comparison. A combined soil–foliar treatment was also conducted to assess the potential synergistic effect of dual nutrient delivery pathways. In this approach, plants received soil injections of MgO-HNTs together with foliar sprays at corresponding concentrations of 0%,

1%, 3%, 5%, and 10%. Four treatment groups were therefore established: untreated controls, soil injection with pure MgO, soil injection with MgO-HNTs, and combined soil and foliar application of MgO-HNTs. All plants were maintained under identical growth conditions. Images were collected on Day 1 and Day 7 to document treatment effects, and both qualitative and quantitative growth outcomes are presented in the Results section.

### 2.7. Measurement of Physiological Growth Indices

Response Surface Methodology (RSM) was employed to optimize seed germination and early seedling development by modeling the combined effects of three independent variables: temperature, MgO-HNT concentration, and light intensity. A total of 17 experimental runs were generated using a Box–Behnken Design (BBD) in Design Expert Software (Version 7), with all experiments conducted in triplicate. A quadratic polynomial model was fitted to evaluate linear, quadratic, and interaction effects among the factors. Seedlings were evaluated on Day 7, and root and shoot lengths were measured using digital calipers and calibrated images analyzed in ImageJ. Eight physiological response variables ($Y_1$–$Y_8$) were quantified to assess germination performance, growth vigor, and stress tolerance: germination percentage ($Y_1$), root length ($Y_2$), shoot length ($Y_3$), total seedling length ($Y_4$), root-to-shoot ratio ($Y_5$), seedling vigor index (SVI; $Y_6$), shoot length stress tolerance index (SLSI; $Y_7$), and root length stress tolerance index (RLSI; $Y_8$).

Germination percentage (GP) was calculated as:

$$\text{GP (\%)} = \left(\frac{n}{N}\right) \times 100 \qquad \text{Eq. 2-1}$$

where $n$ is the number of germinated seeds and $N$ is the total number of seeds sown.

Total seedling length was obtained as:

$$\text{Seedling Length (cm)} = \text{Average Root Length} + \text{Average Shoot Length}$$

The root-to-shoot ratio was calculated as:

$$\text{Root/Shoot Ratio} = \frac{\text{Average Root Length}}{\text{Average Shoot Length}}$$

Seedling vigor index (SVI) was computed using:

$$\text{SVI} = (\text{Average Root Length} + \text{Average Shoot Length}) \times \text{Germination Percentage}$$

Stress tolerance indices were calculated relative to untreated controls:

$$\text{SLSI (\%)} = \left(\frac{\text{Average Shoot Length}_{\text{Treated}}}{\text{Average Shoot Length}_{\text{Control}}}\right) \times 100$$

$$\text{RLSI (\%)} = \left(\frac{\text{Average Root Length}_{\text{Treated}}}{\text{Average Root Length}_{\text{Control}}}\right) \times 100$$

### 2.8. Statistical Analysis

Response Surface Methodology (RSM) was used to analyze the combined effects of temperature, MgO-HNT concentration, and light exposure on seed germination and early seedling growth. A Box–Behnken Design (BBD) with 17 experimental runs (each in triplicate) was generated in Design Expert Software (Version 7). The three independent variables were temperature ($X_1$), MgO-HNT concentration ($X_2$), and light exposure duration ($X_3$). Among the physiological responses measured, seedling length ($Y_4$) and the root length stress tolerance index (RLSI, $Y_8$) were selected for detailed modeling because they provided sensitive indicators of plant vigor and stress resilience. Factor levels were evaluated at three coded values (−1, 0, +1), and the corresponding actual values are listed in Table 2. Quadratic polynomial models were fitted, and model adequacy was assessed using analysis of variance (ANOVA) and diagnostic residual analysis.

**Table 2**: Coded and Actual Levels of Independent Variables Used in RSM

| Variables | Codes | Units | -1 | 0 | +1 |
|---|---|---|---|---|---|
| **Temperature** | $X_1$ | °C | 25 | 35 | 45 |
| **MgO-HNTs Concentration** | $X_2$ | mg | 1 | 50 | 100 |
| **Light** | $X_3$ | h | 8 | 12 | 18 |

### 2.9. Experimental Design

A three-factor experimental design was implemented to evaluate the effects of temperature ($X_1$), MgO-HNT concentration ($X_2$), and light exposure duration ($X_3$) on seed germination and early seedling development. Seventeen experimental runs were conducted using different combinations of the three factors, as summarized in Table 3. Eight physiological responses were measured for each run: germination percentage (%), root length (cm), shoot length (cm), total seedling length (cm), root-to-shoot ratio, seedling vigor index (SVI), shoot length stress tolerance index (SLSI), and root length stress tolerance index (RLSI). These response variables provided the dataset for subsequent ANOVA and Response Surface Methodology (RSM) modeling to determine the statistical significance of each factor and identify optimal germination and growth conditions.

**Table 3:** Experimental Runs and Measured Responses for Seed Germination and Seedling Growth. Temperature ($X_1$), MgO-HNTs concentration ($X_2$), and light exposure ($X_3$) applied in each run, along with the corresponding responses: $Y_1$ (germination percentage yield, %), $Y_2$ (root length, cm), $Y_3$ (shoot length, cm), $Y_4$ (seedling length, cm), $Y_5$ (root/shoot ratio), $Y_6$ (seed vigor index, SVI, %), $Y_7$ (shoot length stress index, SLSI, %), and $Y_8$ (root length stress index, RLSI, %).

| Run | X1 | X2 | X3 | Y1 | Y2 | Y3 | Y4 | Y5 | Y6 | Y7 | Y8 |
|---|---|---|---|---|---|---|---|---|---|---|---|
| 1 | 45 | 50 | 18 | 50 | 11.87 | 6.75 | 18.62 | 1.81 | 9.31 | 165.66 | 100.3 |
| 2 | 25 | 1 | 12 | 75 | 14.56 | 4.53 | 19.09 | 3.18 | 14.31 | 105.14 | 123.03 |
| 3 | 35 | 1 | 18 | 100 | 10.82 | 5.48 | 16.3 | 2.03 | 16.3 | 127.19 | 91.42 |
| 4 | 25 | 50 | 8 | 75 | 10.25 | 5.1 | 15.35 | 2.09 | 11.51 | 118.37 | 86.61 |
| 5 | 35 | 50 | 12 | 100 | 15.92 | 6.01 | 21.93 | 2.64 | 21.93 | 139.49 | 134.52 |
| 6 | 45 | 100 | 12 | 100 | 17.02 | 4.9 | 21.92 | 3.49 | 21.94 | 114.19 | 143.81 |
| 7 | 35 | 1 | 8 | 50 | 8.63 | 4.23 | 12.86 | 2.06 | 6.43 | 98.17 | 72.92 |
| 8 | 35 | 50 | 12 | 75 | 15.23 | 5.59 | 20.82 | 2.75 | 15.61 | 129.74 | 128.69 |
| 9 | 45 | 1 | 12 | 75 | 14.71 | 4.24 | 18.95 | 3.52 | 14.21 | 98.41 | 124.29 |
| 10 | 45 | 50 | 8 | 25 | 8.91 | 4.37 | 13.28 | 2.06 | 3.32 | 101.42 | 75.28 |
| 11 | 35 | 100 | 18 | 75 | 13.73 | 6.12 | 19.85 | 2.26 | 14.88 | 142.04 | 116.01 |

**Table 3:** Experimental Runs and Measured Responses for Seed Germination and Seedling Growth. Temperature ($X_1$), MgO-HNTs concentration ($X_2$), and light exposure ($X_3$) applied in each run, along with the corresponding responses: $Y_1$ (germination percentage yield, %), $Y_2$ (root length, cm), $Y_3$ (shoot length, cm), $Y_4$ (seedling length, cm), $Y_5$ (root/shoot ratio), $Y_6$ (seed vigor index, SVI, %), $Y_7$ (shoot length stress index, SLSI, %), and $Y_8$ (root length stress index, RLSI, %).

| | $X_1$ | $X_2$ | $X_3$ | $Y_1$ | $Y_2$ | $Y_3$ | $Y_4$ | $Y_5$ | $Y_6$ | $Y_7$ | $Y_8$ |
|---|---|---|---|---|---|---|---|---|---|---|---|
| 12 | 35 | 50 | 12 | 100 | 16.04 | 5.1 | 21.14 | 3.22 | 21.14 | 118.37 | 135.53 |
| 13 | 35 | 50 | 12 | 75 | 15.81 | 5.74 | 21.55 | 2.78 | 16.16 | 133.22 | 133.59 |
| 14 | 35 | 50 | 12 | 100 | 16.35 | 5.8 | 22.15 | 2.85 | 22.17 | 135.08 | 138.15 |
| 15 | 25 | 50 | 18 | 75 | 12.36 | 6.41 | 18.77 | 1.94 | 14.07 | 148.77 | 104.44 |
| 16 | 25 | 100 | 12 | 100 | 18.62 | 6.83 | 25.45 | 2.71 | 25.44 | 158.29 | 157.33 |
| 17 | 35 | 100 | 8 | 50 | 9.46 | 4.15 | 13.61 | 2.31 | 6.8 | 96.32 | 79.93 |

## 3. Results and Discussion

### 3.1. Scanning Electron Microscopy (SEM) Analysis

Magnesium oxide (MgO) partially hydrolyzes in aqueous suspension, generating $Mg^{2+}$ ions that exhibit strong electrostatic affinity for the negatively charged siloxane groups on the outer surface of halloysite nanotubes (HNTs). This interaction promotes stable surface modification, while loosely bound MgO particulates are removed during washing. To verify successful deposition, Scanning Electron Microscopy (SEM) and Energy Dispersive Spectroscopy (EDS) analyses were performed. Pristine HNTs (Figure 5A and Figure 5C) displayed smooth, uniform tubular structures with well-defined morphology. Following MgO deposition, the nanotubes exhibited a noticeably rougher and thicker outer surface (Figure 5B and Figure 5D), consistent with the formation of a magnesium-containing layer. Mild agglomeration was also observed, likely resulting from interparticle interactions mediated by surface-bound Mg species. Importantly, the coating remained intact after rinsing under acidic conditions (pH ~ 4), indicating strong and stable $Mg^{2+}$–HNT interactions. Similar acid-resistant binding behavior has been reported for metal-oxide functionalization of halloysite, where divalent cations form durable electrostatic or coordination bonds with the siloxane exterior [49,50]. This stability supports the intended application of MgO-HNTs in soil environments, where pH fluctuations are common and coating durability is essential for nutrient delivery.

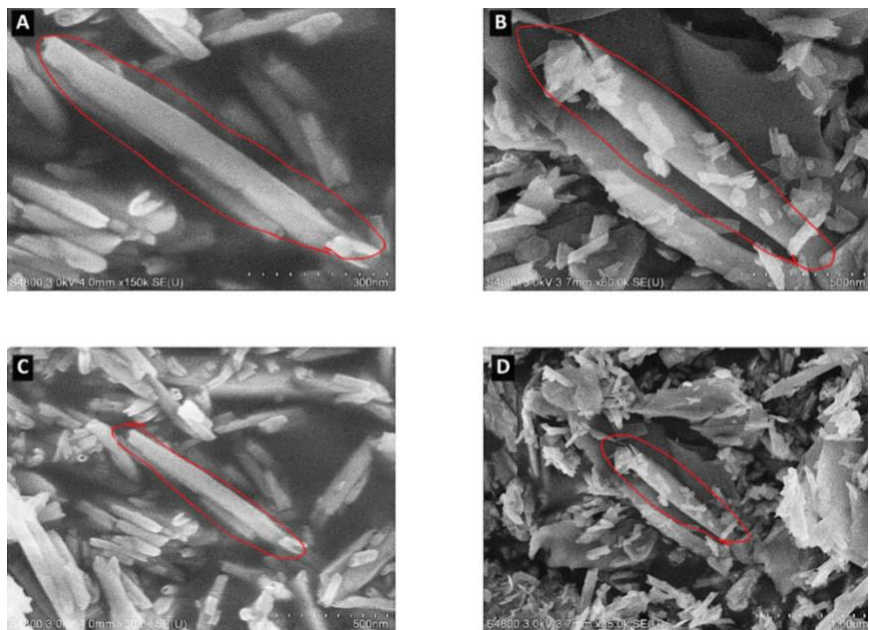

**Figure 5**: SEM micrographs showing morphology and surface structure (A–D). (A) Pristine HNT (300 nm), (B) MgO-coated HNT (500 nm), (C) Pristine HNT (500 nm), and (D) MgO-coated HNT (1.00 μm).

### 3.2. EDS Elemental Analysis
#### 3.2.1. Qualitative Spectral Analysis

Energy-dispersive X-ray spectroscopy (EDS) was used to compare the elemental composition of pristine HNTs and MgO-coated HNTs (Figure 6). The pristine sample (Figure 6A) exhibited the characteristic peaks of halloysite—aluminum (Al), silicon (Si), and oxygen (O)—along with a minor carbon (C) signal attributed to surface contamination. As expected, no magnesium peak was detected. In contrast, the MgO-HNT sample (Figure 6B) displayed a distinct Mg peak at ~1.25 keV, confirming successful deposition of magnesium oxide onto the nanotube surface. The retention of Al, Si, and O peaks indicates preservation of the underlying HNT framework. These results provide direct elemental evidence of Mg incorporation and validate the effectiveness of the coating process. The observed increase in surface roughness and Mg signals in the EDS spectra is consistent with previous reports showing strong electrostatic binding of $Mg^{2+}$ to the negatively charged siloxane surface of halloysite nanotubes. The resulting MgO-HNT architecture enhances water retention and provides localized nutrient release, which are known to support early seedling vigor in soil-limited environments. Similar coating behavior and improved plant responses have been documented for metal-oxide–modified nanoclays used in agriculture [49,50].

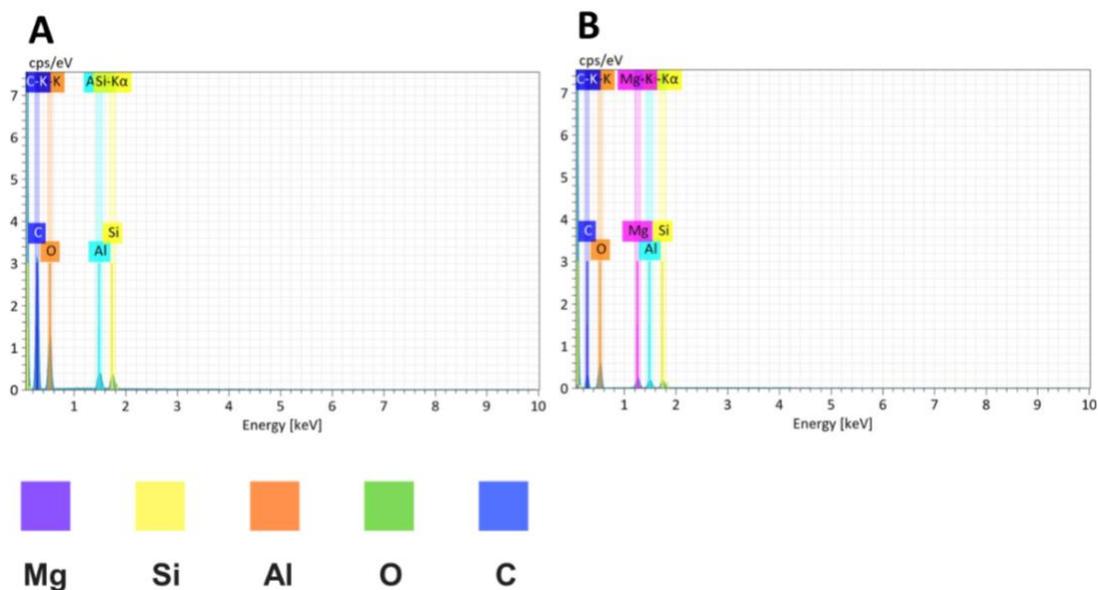

**Figure 6:** EDS spectra of (A) pristine HNTs showing no Mg signal and (B) MgO-HNTs exhibiting a clear Mg peak, confirming surface modification. Characteristic HNT elements (Al, Si, O) and trace carbon (C) are present in both samples.

### 3.2.2. Quantitative Elemental Composition

Quantitative EDS analysis confirmed the successful incorporation of magnesium onto the halloysite nanotube surface (Tables 4 and 5). Pristine HNTs displayed the expected aluminosilicate signature—oxygen, aluminum, and silicon—consistent with the known composition of halloysite nanotubes [1]. No magnesium was detected in the unmodified sample. In contrast, MgO-HNTs exhibited a clear magnesium peak (6.03 atom %), verifying effective MgO deposition. Minor shifts in oxygen and carbon content reflect surface chemical changes associated with the coating process. The continued presence of aluminum and silicon indicates preservation of the underlying halloysite structure [49]. These findings align with previous reports showing strong electrostatic interactions between $Mg^{2+}$ ions and the negatively charged siloxane groups on the HNT surface, enabling stable metal-oxide attachment [50]. The successful integration confirmed by EDS supports the use of MgO-HNTs as a functional nanomaterial for improving nutrient retention and early seedling performance.

**Table 4**: EDS Elemental Analysis of Pristine Halloysite Nanotubes (HNTs)

| Element | At. No. | Netto | Mass [%] | Mass Norm. [%] | Atom [%] | Abs. error [%] (1 sigma) | Rel. error [%] (1 sigma) |
|---|---|---|---|---|---|---|---|
| **Carbon** | 6 | 34176 | 52.20 | 50.93 | 63.15 | 6.37 | 12.21 |
| **Oxygen** | 8 | 15106 | 27.30 | 26.64 | 24.80 | 3.64 | 13.35 |
| **Aluminium** | 13 | 5734 | 7.23 | 7.05 | 3.89 | 0.40 | 5.50 |
| **Silicon** | 14 | 5580 | 15.77 | 15.38 | 8.16 | 0.82 | 5.19 |

**Table 5**: EDS Elemental Analysis of MgO-Coated Halloysite Nanotubes (MgO-HNTs)

| Element | At. No. | Netto | Mass [%] | Mass Norm. [%] | Atom [%] | Abs. error [%] (1 sigma) | Rel. error [%] (1 sigma) |
|---|---|---|---|---|---|---|---|
| **Carbon** | 6 | 3388 | 21.04 | 30.11 | 40.35 | 3.58 | 16.99 |
| **Oxygen** | 8 | 7148 | 30.04 | 43.00 | 43.25 | 4.45 | 14.81 |
| **Magnesium** | 12 | 3165 | 6.36 | 9.11 | 6.03 | 0.39 | 6.12 |
| **Aluminium** | 13 | 2678 | 5.33 | 7.62 | 4.55 | 0.30 | 5.72 |
| **Silicon** | 14 | 2830 | 7.09 | 10.15 | 5.82 | 0.37 | 5.16 |

### 3.3. Seed Germination Performance Under Different Treatments

Figure 7 illustrates the progression of tomato seedlings subjected to foliar spraying with MgO-HNTs compared with untreated controls. Over the 7-day period, the control plants ($S_1$–$S_2$) exhibited only modest increases in leaf size and height, indicating limited natural growth under baseline conditions. In contrast, seedlings receiving foliar MgO-HNT treatments ($S_4$–$S_6$) showed visibly enhanced vigor. Before treatment, seedlings in $S_3$ and $S_5$ appeared less robust, but following foliar application, the corresponding samples ($S_4$ and $S_6$) demonstrated marked improvements in leaf expansion, stem rigidity, and overall plant vitality. These improvements align with reports showing that foliar-applied Mg-based nanomaterials enhance stomatal penetration and chlorophyll biosynthesis by improving magnesium bioavailability [51,52].

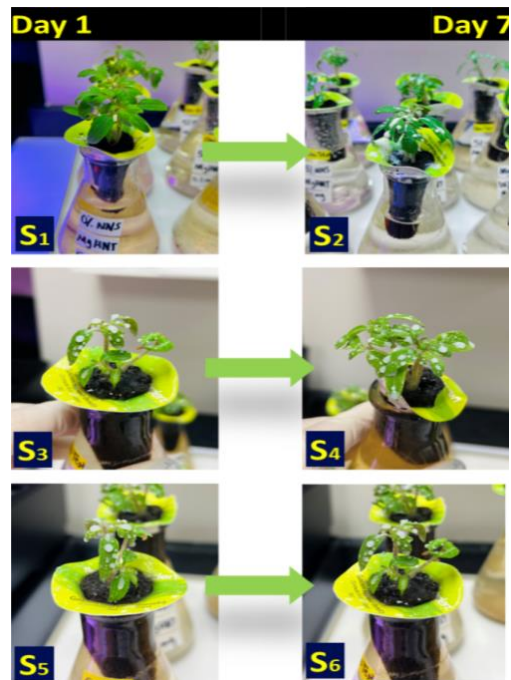

**Figure 7:** Growth Progression of Tomato Seedlings Before and After MgO-HNTs Foliar Application Compared to Control. $S_1$–$S_2$: Untreated control group. $S_4$–$S_6$: Treated groups — seedlings before **($S_3$, $S_5$)** and after **($S_4$, $S_6$)** MgO-HNTs foliar application.

The independent evaluation of soil-based nutrient delivery is shown in Figure 8. Seedlings receiving higher MgO-HNT dosages exhibited more rapid leaf expansion and stronger early biomass accumulation. This dose-responsive improvement is consistent with halloysite nanotubes' ability to retain moisture and provide slow, sustained nutrient release within the rhizosphere, thereby supporting early root establishment [53, 54].

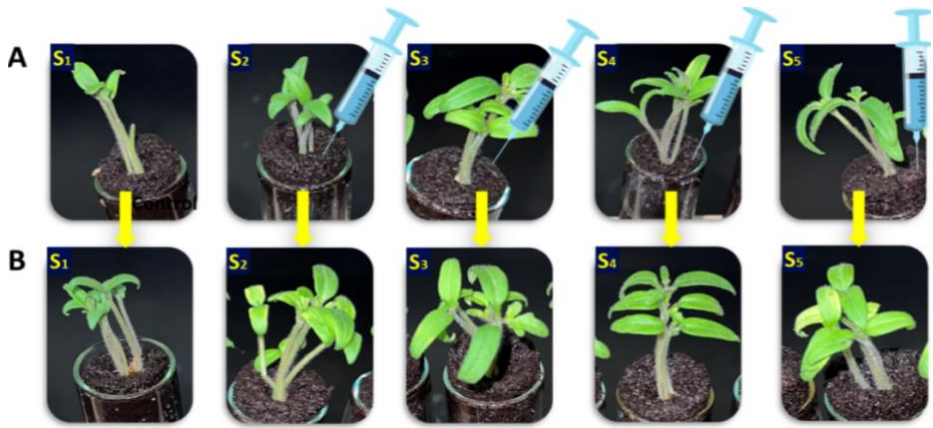

**Figure 8:** Soil injection effects of MgO-HNTs on tomato seedling growth. Tomato seedlings were treated with five MgO-HNT concentrations ($S_1$: 0.1 mg/mL, $S_2$: 1 mg/mL, $S_3$: 10 mg/mL, $S_4$: 50 mg/mL, $S_5$: 100 mg/mL). Panels show seedling growth at Day 1 (A) and Day 7 (B), demonstrating a dose-dependent improvement in early development.

Figure 9 presents the final phase of experimentation, in which soil injections and foliar sprays were combined. Across all groups, MgO-HNT–treated seedlings outperformed both pure-MgO treatments and controls, with the combined soil–foliar application producing the strongest responses. Similar synergistic effects have been observed when nanocarrier-delivered nutrients are supplied through both foliar and root pathways, resulting in enhanced uptake efficiency and reduced early nutrient stress [55].

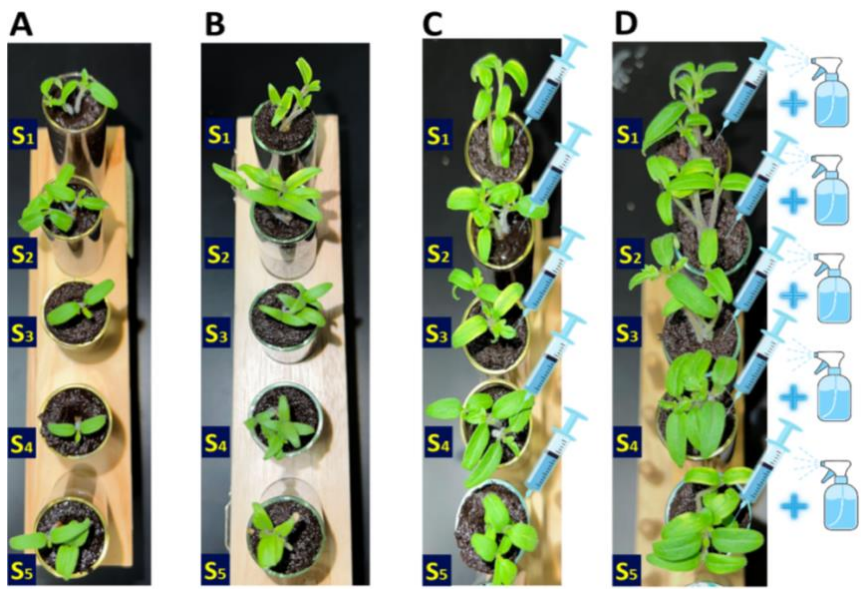

**Figure 9:** Effects of MgO and MgO-HNT soil and soil–foliar treatments on tomato seedling growth. (A) Control; (B) soil injection with pure MgO; (C) soil injection with MgO-HNTs; (D) combined MgO-HNT soil injection with foliar sprays (0–10%). Treatments correspond to five MgO-HNT concentrations: $S_1$ = 0.1 mg/mL, $S_2$ = 1 mg/mL, $S_3$ = 10 mg/mL, $S_4$ = 50 mg/mL, and $S_5$ = 100 mg/mL

Root development responses are shown in Figure 10 for both heirloom cherry and golden tomato varieties. Control seedlings displayed limited root elongation at Day 7. By Day 14, seedlings exposed to intermediate and high MgO-HNT concentrations—especially 10 mg and 100 mg—exhibited substantially greater root length and branching. This enhanced root development is consistent with previous findings showing that nanoclay carriers improve water retention, moderate oxidative stress, and maintain nutrient microzones around developing roots, promoting deeper and more extensive root systems [53, 56].

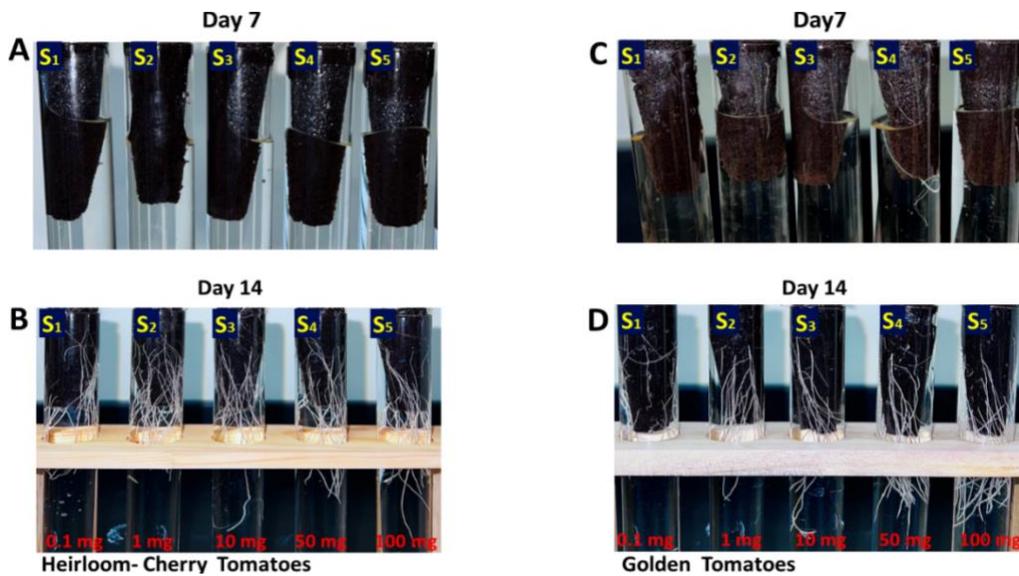

**Figure 10:** (A-D) Root development of heirloom cherry and golden tomato seedlings treated with MgO-HNTs at five concentrations ($S_1$ = 0.1 mg/mL, $S_2$ = 1 mg/mL, $S_3$ = 10 mg/mL, $S_4$ = 50 mg/mL, $S_5$ = 100 mg/mL) at Day 7 and Day 14.

### 3.4. Seedling Vigor and Shoot Length Stress Tolerance under Different Treatments

The effects of temperature, MgO-HNT concentration, and light exposure on seedling vigor and shoot length stress tolerance were evaluated across 17 experimental runs. The Seedling Vigor Index (SVI, %) for each run is shown in Figure 11A, where values ranged from 3.32% to 25.44%. The highest vigor was observed in Run 16 (25.44%), corresponding to moderate temperature, high MgO-HNT concentration, and a 12-hour photoperiod. The lowest SVI occurred in Run 10 (3.32%), indicating unfavorable environmental and treatment conditions for promoting early seedling growth. The Shoot Length Stress Index (SLSI, %) is presented in

Figure 11B, with values ranging from 96.32% to 165.66%. Run 1 exhibited the highest SLSI (165.66%), associated with elevated temperature and moderate MgO-HNT supplementation under extended light exposure. In contrast, Run 17 produced the lowest SLSI (96.32%), reflecting reduced shoot growth adaptability under that combination of factors. These trends demonstrate that MgO-HNT supplementation enhances metabolic activity and stress buffering when combined with optimal temperature and light exposure. Magnesium is known to improve chlorophyll synthesis and early growth vigor, while nanoclay carriers improve water retention and nutrient accessibility, supporting greater resilience under fluctuating light or temperature stress [51,53]. Accordingly, SVI and SLSI provide reliable indicators for subsequent RSM optimization of early plant performance.

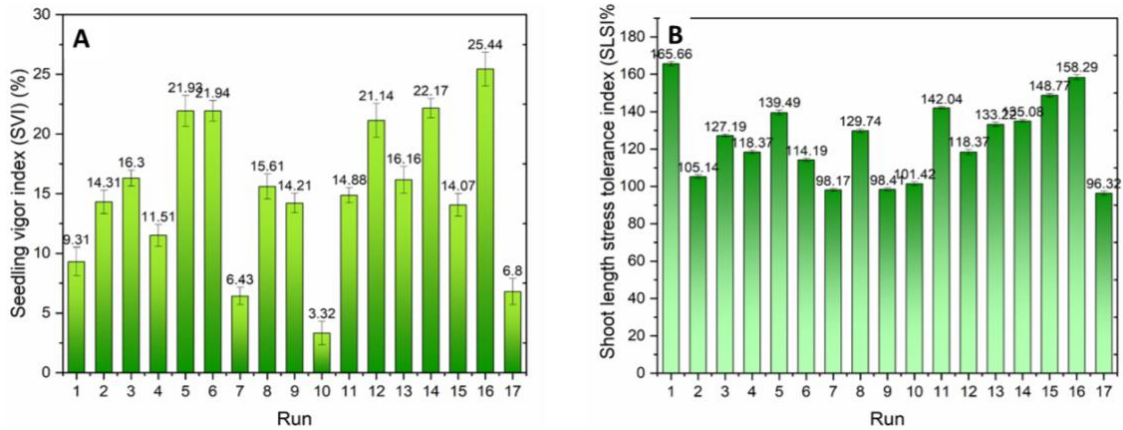

**Figure 11**: Effects of Different Treatments on Seedling Vigor and Shoot Length Stress Tolerance. (A) Seedling Vigor Index (SVI, %) across 17 experimental runs. (B) Shoot Length Stress Index (SLSI, %) across 17 experimental runs.

### 3.5. ANOVA and Model Results

Preliminary response trends indicated that Seedling Length ($Y_4$) and Root Length Stress Tolerance Index (RLSI, $Y_8$) were the most biologically sensitive to treatment conditions; therefore, these indices were selected for detailed statistical modeling. ANOVA was conducted to evaluate the significance of temperature ($X_1$), MgO-HNTs concentration ($X_2$), and light exposure ($X_3$) on these responses and to assess the reliability of the fitted models. The full ANOVA results for $Y_4$ and $Y_8$ are shown in Table 2-5 and Table 2-6, summarizing the sum of squares, degrees of freedom, mean square, F-values, and p-values for each factor and interaction. For seedling length (Table 6), the model was highly significant ($p < 0.0001$), confirming strong predictive capability. $X_2$ (MgO-HNTs concentration) and $X_3$ (light exposure) both had highly significant effects ($p < 0.0001$), and the quadratic term $X_3^2$ was also significant ($p < 0.0001$), indicating a nonlinear light response. $X_1$ (temperature), $X_1^2$, and $X_2^2$ were not significant within the tested range. The lack-of-fit was non-significant ($p = 0.1735$), confirming an adequate model fit. These results show that seedling elongation is primarily controlled by MgO-HNT dosage and photoperiod, whereas temperature contributes minimally under the tested conditions. For RLSI (Table 7), the model was again highly significant (F = 56.2, p < 0.0001). $X_2$ (MgO-HNTs concentration) and $X_3$ (light) significantly affected root stress tolerance (p = 0.0003 and p = 0.0001, respectively), with $X_3^2$ showing a strong nonlinear effect (p < 0.0001). Temperature ($X_1$) approached significance (p = 0.0838) but was not statistically meaningful, and neither $X_1^2$ nor $X_2^2$ showed significant contributions. The lack-of-fit test was non-significant (p = 0.1735), confirming good model adequacy. These findings indicate that MgO-HNT concentration and controlled light exposure are key drivers of root stress resilience in early seedlings.

**Table 6:** ANOVA table for Seedling Length Yield ($Y_4$). The model was statistically significant (p < 0.05), with MgO-HNTs concentration ($X_2$), light exposure ($X_3$), and the quadratic effect of light ($X_3^2$) showing significant impacts. The non-significant lack-of-fit confirms the model's good fit with the observed data.

| Source | Sum of Squares | df | Mean Square | F Value | p-value Prob > F | P value | Significance |
|---|---|---|---|---|---|---|---|
| **Model** | 198.67 | 9 | 22.07 | 29.37 | < 0.0001 | < 0.05 | significant |
| **$X_1$-Temperature** | 3.74 | 1 | 3.74 | 4.97 | 0.061 | | |
| **$X_2$- MgO-HNTs** | 24.31 | 1 | 24.31 | 32.35 | 0.0007 | < 0.05 | significant |
| **$X_3$-Light** | 42.61 | 1 | 42.61 | 56.69 | 0.0001 | < 0.05 | significant |
| **$X_1^2$** | 0.49 | 1 | 0.49 | 0.66 | 0.4438 | | |
| **$X_2^2$** | 1.16 | 1 | 1.16 | 1.54 | 0.254 | | |
| **$X_3^2$** | 139.5 | 1 | 139.5 | 185.62 | < 0.0001 | < 0.05 | significant |
| **Residual** | 5.26 | 7 | 0.75 | | | | |
| **Lack of Fit** | 4.06 | 3 | 1.35 | 4.51 | 0.0898 | | not significant |
| **Pure Error** | 1.2 | 4 | 0.3 | | | < 10 | |
| **Cor Total** | 203.94 | 16 | | | | | |

**Table 7:** ANOVA table for Root Length Stress Tolerance Index Yield ($Y_8$). Significant effects were observed for MgO-HNTs concentration ($X_2$), light ($X_3$), and its quadratic term ($X_3^2$), indicating strong influence on root stress resilience. The model passed the lack-of-fit test, supporting its predictive reliability.

| Source | Sum of Squares | df | Mean Square | F Value | p-value Prob > F | P value | Significance |
|---|---|---|---|---|---|---|---|
| **Model** | 10759.19 | 9 | 1195.47 | 56.2 | < 0.0001 | < 0.05 | significant |
| **$X_1$- Temperature** | 86.32 | 1 | 86.32 | 4.06 | 0.0838 | | |
| **$X_2$- MgO-HNTs** | 958.96 | 1 | 958.96 | 45.08 | 0.0003 | < 0.05 | significant |
| **$X_3$-Light** | 1190.24 | 1 | 1190.24 | 55.96 | 0.0001 | < 0.05 | significant |
| **$X_1^2$** | 22.41 | 1 | 22.41 | 1.05 | 0.3388 | | |
| **$X_2^2$** | 1.56 | 1 | 1.56 | 0.073 | 0.7946 | | |
| **$X_3^2$** | 9178.21 | 1 | 9178.21 | 431.49 | < 0.0001 | < 0.05 | significant |
| **Residual** | 148.9 | 7 | 21.27 | | | | |
| **Lack of Fit** | 100.74 | 3 | 33.58 | 2.79 | 0.1735 | | not significant |
| **Pure Error** | 48.15 | 4 | 12.04 | | | | |
| **Cor Total** | 10908.09 | 16 | | | | | |

A fundamental assumption in Response Surface Methodology (RSM) is that the residuals of the model are normally distributed [57, 58]. To evaluate this, normal probability plots were generated for both response variables: Seedling Length ($Y_4$) and Root Length Stress Tolerance Index ($Y_8$). These plots visually represent the distribution of residuals compared to a theoretical normal distribution. As illustrated in Figure 12, the residuals for both models show a strong linear alignment along the red reference line. This pattern suggests that the residuals are symmetrically distributed and closely approximate a normal distribution. The absence of significant deviations or curvature in the plots indicates that the models meet the assumption of residual normality, thereby supporting the statistical reliability and validity of the RSM-based predictions. The normality of residuals is particularly important in regression modeling because it ensures the validity of hypothesis testing and confidence intervals derived from the models. The observed linear trends confirm that the experimental data are well-suited for RSM analysis and that the models can be used for accurate prediction and optimization. The close alignment of data points with the reference line indicates approximate normality of residuals, validating the assumptions underlying the RSM analysis.

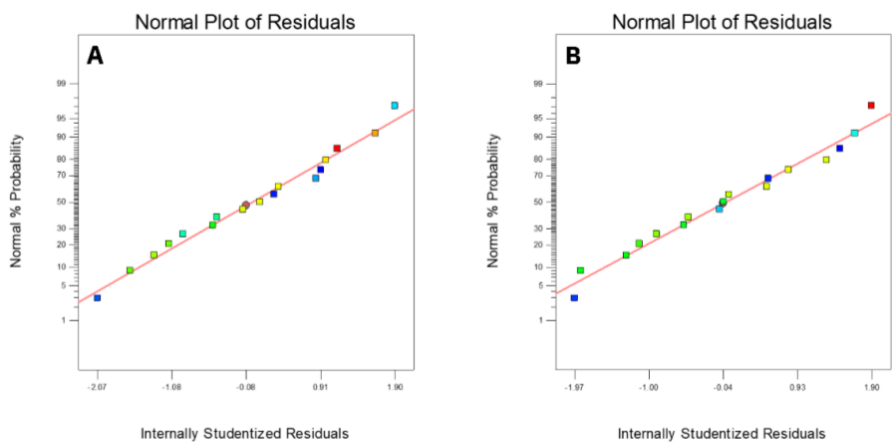

**Figure 12:** Normal probability plots of residuals for RSM model diagnostics. A) Root Length Stress Tolerance Index ($Y_8$); B) Seedling Length ($Y_4$).

### 3.6. Response Surface Analysis of Seedling Length ($Y_4$)

The 3D response surfaces illustrate how temperature ($X_1$), MgO-HNT concentration ($X_2$), and light exposure ($X_3$) interact to regulate early tomato seedling elongation. Each surface represents a pairwise interaction while the third factor is held at its midpoint. Temperature and light exposure jointly influence elongation in Figure 13A, where maximum seedling length occurs under low temperature and high—but not extreme—light levels. The curvature along the light axis reflects that photosynthetic gain is limited by the plant's stress threshold. This balance between reduced metabolic demand at cooler temperatures and increased carbon assimilation under adequate light aligns with established models of early seedling growth physiology [59]. The interaction between MgO-HNT concentration and temperature (Figure 13B) demonstrates that the longest seedlings occur at higher MgO-HNT levels combined with lower temperatures. Cooler temperatures decrease respiratory losses and slow water evaporation, allowing plants to more efficiently use the nutrient-rich microenvironment created by MgO-HNTs. Similar temperature–nanomaterial synergies have been observed in studies where metal-oxide nanostructures mitigate early-stage abiotic stress by stabilizing the rhizosphere [60]. Seedling length responds nonlinearly to the interaction between MgO-HNT concentration and light (Figure 13C). Increasing MgO-HNT levels enhances elongation under moderate light, likely due to improved nutrient availability and moisture retention around the root zone. However, at high light intensities, growth plateaus, consistent with reduced photosynthetic efficiency under excessive irradiance (photoinhibition) [61]. This trend underscores that MgO-HNT benefits are maximized only when photoperiod is regulated to avoid light-induced stress.

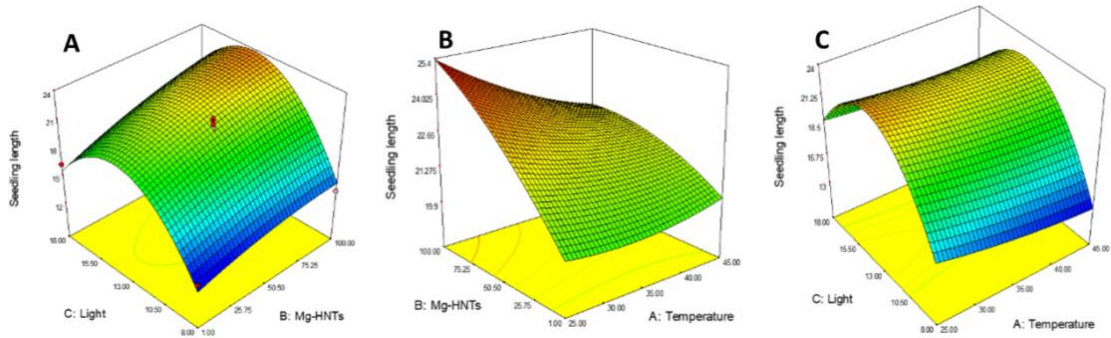

**Figure 13:** Response surface plots illustrating interaction effects of (A) Temperature ($X_1$) and Light ($X_3$), (B) MgO-HNTs concentration ($X_2$) and Temperature ($X_1$), and (C) MgO-HNTs concentration ($X_2$) and Light ($X_3$) on Seedling Length ($Y_4$).

### 3.7. Response Surface Analysis of Root Length Stress Tolerance (RLSI, $Y_8$)

Response surface methodology was applied to evaluate how temperature ($X_1$), MgO-HNT concentration ($X_2$), and light exposure ($X_3$) jointly influence the Root Length Stress Tolerance Index (RLSI, $Y_8$). The three-dimensional plots in Figure 14 illustrate these interaction effects. The interaction between MgO-HNT concentration and light (Figure 14A) exhibits a curved response, where RLSI increases at higher MgO-HNT levels under low to moderate light. This pattern suggests that nanomaterial-enhanced nutrient and moisture availability improves root tolerance. At high light intensities, however, the benefit plateaus, consistent with reduced photosynthetic efficiency or light-induced metabolic stress under excessive irradiance (photoinhibition) [61]. The MgO-HNT concentration–temperature interaction (Figure 14B) shows that RLSI decreases sharply as temperature approaches 45 °C, even at high nanomaterial levels. This decline aligns with known thermal limits in tomato seedlings, where temperatures above ~35 °C impair membrane stability, water uptake, and root metabolic activity [62]. Optimal

RLSI occurs at moderate temperatures (25–30 °C) combined with elevated MgO-HNT concentrations. The temperature–light interaction (Figure 14C) indicates that RLSI is maximized under moderate temperatures with increased light exposure. Cooler conditions help reduce stress respiration, while adequate light supports energy production for root recovery and elongation. The curvature of the surface suggests that both factors must be balanced, as extremes in either variable diminish overall stress tolerance.

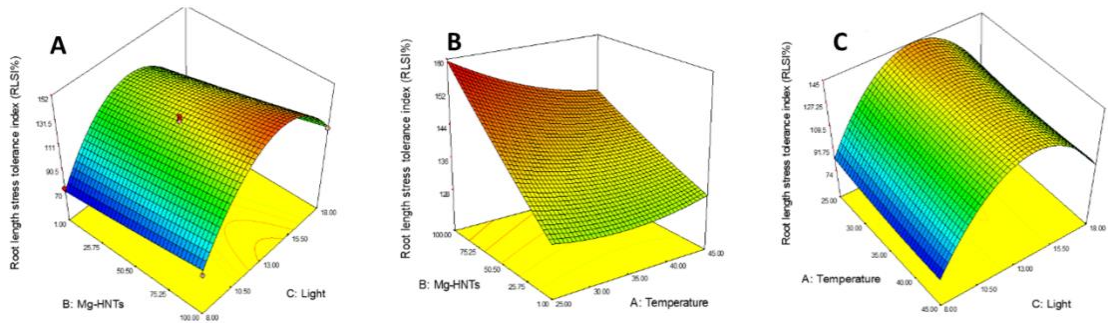

**Figure 14:** Response surface plots illustrating interaction effects of (A) MgO-HNTs concentration ($X_2$) and Light ($X_3$), (B) MgO-HNTs concentration ($X_2$) and Temperature ($X_1$), and (C) Temperature ($X_1$) and Light ($X_3$) on Root Length Stress Tolerance Index (RLSI%, $Y_8$).

### 3.8. Comparative Root Development in Lunar and Martian Soil Simulants

After determining the optimal RSM growth conditions (25 °C and a 12 h photoperiod), these parameters were applied to extraterrestrial soil simulants to evaluate seedling performance under resource-limited environments. Five concentrations of MgO-HNTs (0.1, 1, 10, 50, and 100 mg) were incorporated into both lunar and Martian regolith simulants (Figure 15). Heirloom Cherry Tomato seeds were used to maintain consistency with previous assays, and quantitative measurements of root length, shoot length, and germination percentage were obtained using ImageJ. In the lunar regolith simulant, root development increased progressively with MgO-HNT concentration. The strongest enhancement occurred at 100 mg, where root length reached 17.704 mm, shoot length 5.08 mm, and germination 80%. These improvements reflect the ability of halloysite nanotubes to compensate for the regolith's poor water-holding capacity and minimal nutrient availability. The hollow tubular structure of HNTs can act as a microscale reservoir that retains moisture and provides slow ion release, thereby reducing abiotic stress and supporting sustained root expansion [63]. In contrast, the Martian regolith simulant exhibited a different dose–response pattern. Maximum root length (12.284 mm) occurred at the lower dose of 10 mg MgO-HNTs. Although shoot length and germination continued to rise at higher concentrations—reaching 4.276 mm and 100% at 100 mg—root elongation declined slightly beyond the optimum. This divergence likely reflects interactions between elevated nanoparticle concentrations and the regolith's iron-rich, oxidizing chemistry. Martian soil analogs are known to impose oxidative stress, disrupt root membrane stability, and alter rhizosphere pH, which can collectively inhibit elongation at higher amendment levels [64, 65]. These results demonstrate that optimal MgO-HNT dosing is strongly soil-type dependent. Lunar simulants respond favorably to higher MgO-HNT concentrations due to their extreme nutrient and moisture limitations, whereas Martian simulants achieve maximum root performance at substantially lower doses. This distinction has direct relevance for in-situ resource utilization (ISRU) and extraterrestrial agriculture, where minimizing amendment quantities while maintaining robust early seedling establishment is essential for long-term sustainability.

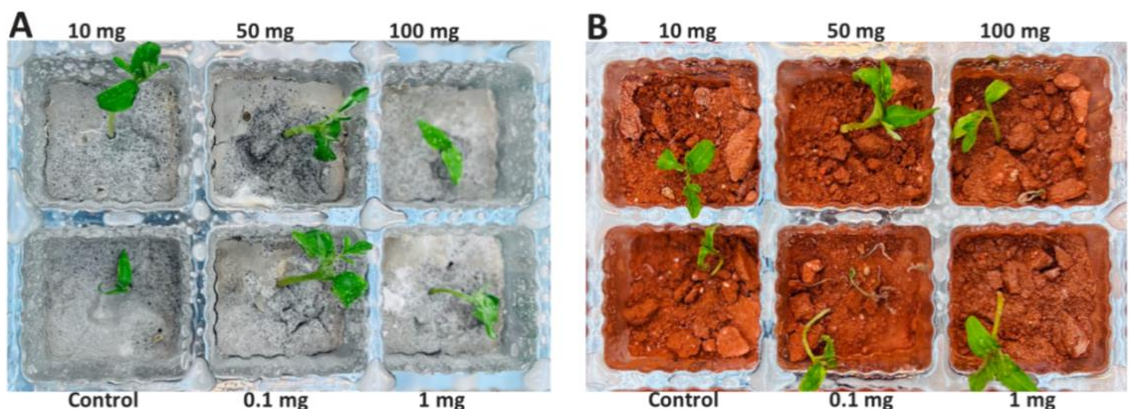

**Figure 15.** Root development in extraterrestrial soil simulants. (A) Lunar regolith showing maximal root length at 100 mg/L MgO-HNTs. (B) Martian regolith showing peak root length at 10 mg/L MgO-HNTs.

To assess the early morphological responses of tomato seedlings to MgO-HNT treatments in extraterrestrial soil simulants, plants grown in lunar and Martian regolith were photographed after 14 days (Figure 16). Seedlings were exposed to six MgO-HNT concentrations (0, 0.1, 1, 10, 50, and 100 mg), corresponding to treatments $S_1$–$S_6$. Panels (A) and (B) show seedlings grown in lunar and Martian regolith, respectively, with red dashed lines serving as reference markers for comparing shoot height and root extension. Root lengths were quantified using ImageJ. Across both soil types, seedlings exhibited improved shoot and root development at intermediate MgO-HNT concentrations, indicating that the nanomaterial enhances early growth by improving moisture retention and localized nutrient availability in nutrient-poor regolith substrates [63]. The slightly different responses between lunar and Martian soils likely reflect variation in their geochemistry, particularly the oxidizing, iron-rich nature of Martian regolith, which can influence root physiology and stress responses [65].

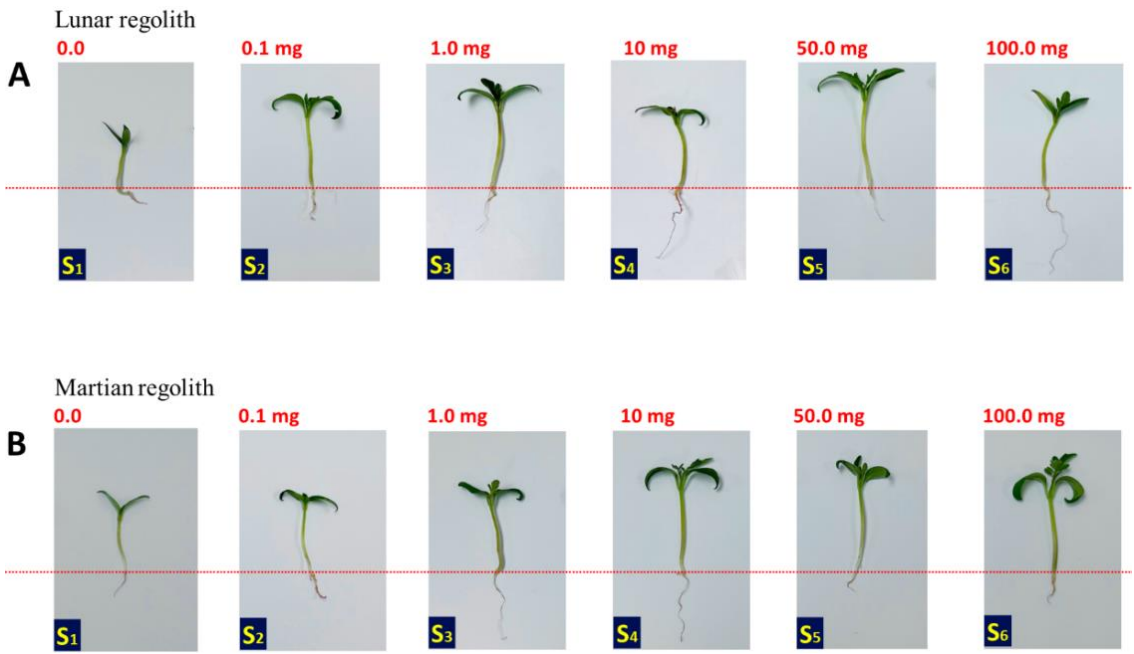

**Figure 16**: Tomato seedling morphology after 14 days in extraterrestrial soil simulants. (A) Lunar regolith. (B) Martian regolith. Seedlings were treated with MgO-HNTs at 0, 0.1, 1.0, 10, 50, and 100 mg. Red dashed lines indicate reference heights for shoot and root comparison. Root measurements obtained via ImageJ.

The effects of MgO-HNTs concentration on root development in extraterrestrial soil simulants are shown in Figure 17. Root lengths were quantified using ImageJ to provide objective assessment of morphological responses. Panel (A) presents root growth in the lunar regolith simulant under a 12-hour photoperiod, while panel (B) shows corresponding measurements for seedlings grown in the Martian regolith simulant under the same conditions. Root elongation varied with both soil type and MgO-HNTs concentration: seedlings in the lunar simulant exhibited greater overall root length compared to those in the Martian simulant. These results highlight how soil composition, light exposure, and MgO-HNT supplementation jointly influence early root development in simulated space-agriculture environments.

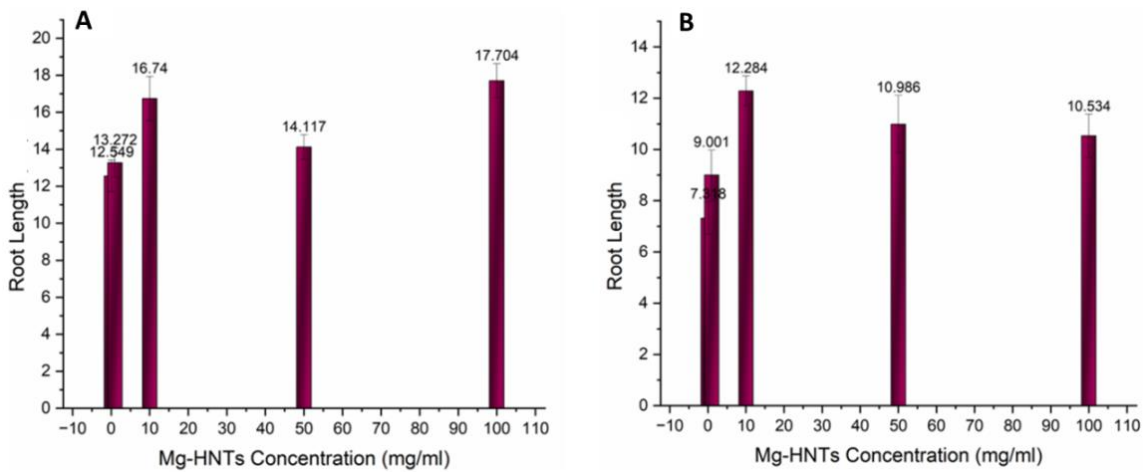

**Figure 17:** Effect of MgO-HNT concentration on root length in extraterrestrial soil simulants. (A) Lunar regolith. (B) Martian regolith. Error bars represent standard deviation. MgO-HNT concentrations: 0–100 mg/mL. Root lengths quantified using ImageJ.

## 4. Conclusion

This research presented the synthesis, characterization, and application of magnesium oxide–coated halloysite nanotubes (MgO-HNTs) as an emerging nanomaterial to improve seed germination and early tomato seedling development. MgO-HNTs were fabricated through an electrodeposition method, and SEM–EDS analyses confirmed successful surface coating and preservation of the halloysite tubular framework. Plant growth experiments were evaluated using a 17-run Box–Behnken response surface design incorporating temperature ($X_1$), MgO-HNT concentration ($X_2$), and photoperiod ($X_3$). Statistical modeling showed high significance and predictive strength. Among eight measured physiological indices, seedling length ($Y_4$) and root length stress tolerance index (RLSI, $Y_8$) emerged as the most responsive variables for optimization. ANOVA, residual diagnostics, and 3D response surfaces collectively showed that MgO-HNT concentration and light exposure were the primary drivers of seedling performance, while temperature played a secondary but supportive role. Optimal growth occurred under 25 °C, a 12-hour light cycle, and 100 mg/L MgO-HNTs, conditions that maximized vigor, elongation, and stress resilience. When these optimized conditions were applied to lunar and Martian regolith simulants, MgO-HNTs substantially improved germination, root penetration, and early biomass accumulation despite the severe nutrient and structural limitations of these substrates. Root development peaked at 100 mg/L in lunar simulant and at 10 mg/L in Martian simulant, reflecting soil-dependent interactions between MgO-HNTs and regolith chemistry. This study demonstrates that MgO-HNTs offer a robust strategy for enhancing water retention, nutrient availability, and physiological stress tolerance in plants. Their effectiveness in both Earth soils and extraterrestrial simulants highlights their potential not only for sustainable terrestrial agriculture but also for future bioregenerative life-support systems designed for long-duration missions to the Moon and Mars.

## 5. Author Contributions

Conceptualization, Z.J.V. and D.K.M.; Methodology, Z.J.V.; Software, Z.J.V.; Validation, Z.J.V. and D.K.M.; Formal Analysis, Z.J.V.; Investigation, Z.J.V.; Writing—Original Draft Preparation, Z.J.V.; Writing—Review and Editing, D.K.M.; Visualization, Z.J.V.; Project Administration, D.K.M.; Funding Acquisition, D.K.M.

## 6. Funding

This research was supported by the NASA EPSCoR Rapid Response Research (R3) Program, Award No. 22-2022-R30015.

## 7. Conflicts of Interest

The authors declare no conflict of interest.

# References

1. Ma C, Yang Z, Xia R, Song J, Liu C, Mao R, Li M, Qin X, Hao C, Jia R. Rising water pressure from global crop production—a 26-yr multiscale analysis. Resources, Conservation and Recycling. 2021 Sep 1;172:105665. https://doi.org/10.1016/j.resconrec.2021.105665

2. Hemathilake DM, Gunathilake DM. Agricultural productivity and food supply to meet increased demands. InFuture foods 2022 Jan 1 (pp. 539-553). Academic Press. https://doi.org/10.1016/B978-0-323-91001-9.00016-5

3. Prăvălie R, Patriche C, Borrelli P, Panagos P, Roșca B, Dumitrașcu M, Nita IA, Săvulescu I, Birsan MV, Bandoc G. Arable lands under the pressure of multiple land degradation processes. A global perspective. Environmental Research. 2021 Mar 1;194:110697. https://doi.org/10.1016/j.envres.2020.110697

4. Al-Mamun MR, Hasan MR, Ahommed MS, Bacchu MS, Ali MR, Khan MZ. Nanofertilizers towards sustainable agriculture and environment. Environmental Technology & Innovation. 2021 Aug 1;23:101658. https://doi.org/10.1016/j.eti.2021.101658

5. Devaux A, Goffart JP, Kromann P, Andrade-Piedra J, Polar V, Hareau G. The potato of the future: opportunities and challenges in sustainable agri-food systems. Potato Research. 2021 Dec;64(4):681-720. https://doi.org/10.1007/s11540-021-09532-x

6. Sarkar MR, Rashid MH, Rahman A, Kafi MA, Hosen MI, Rahman MS, Khan MN. Recent advances in nanomaterials based sustainable agriculture: An overview. Environmental Nanotechnology, Monitoring & Management. 2022 Dec 1;18:100687. https://doi.org/10.1016/j.enmm.2022.100687

7. Garg S, Rumjit NP, Roy S. Smart agriculture and nanotechnology: Technology, challenges, and new perspective. Advanced Agrochem. 2024 Jun 1;3(2):115-25. https://doi.org/10.1016/j.aac.2023.11.001

8. Mortimer JC, Gilliham M. SpaceHort: redesigning plants to support space exploration and on-earth sustainability. Current Opinion in Biotechnology. 2022 Feb 1;73:246-52. https://doi.org/10.1016/j.copbio.2021.08.018

9. An C, Sun C, Li N, Huang B, Jiang J, Shen Y, Wang C, Zhao X, Cui B, Wang C, Li X. Nanomaterials and nanotechnology for the delivery of agrochemicals: strategies towards sustainable agriculture. Journal of Nanobiotechnology. 2022 Jan 4;20(1):11. https://doi.org/10.1186/s12951-021-01214-7

10. Mavridis K, Todas N, Kalompatsios D, Athanasiadis V, Lalas SI. Lycopene and Other Bioactive Compounds' Extraction from Tomato Processing Industry Waste: A Comparison of Ultrasonication Versus a Conventional Stirring Method. Horticulturae. 2025 Jan 10;11(1):71. https://doi.org/10.3390/horticulturae11010071

11. Peixoto FB, Raimundini Aranha AC, Nardino DA, Defendi RO, Suzuki RM. Extraction and encapsulation of bioactive compounds: A review. Journal of Food Process Engineering. 2022 Dec;45(12):e14167. https://doi.org/10.1111/jfpe.14167

12. Gajanan K, Tijare SN. Applications of nanomaterials. Materials Today: Proceedings. 2018 Jan 1;5(1):1093-6. https://doi.org/10.1016/B978-0-12-816806-6.00015-7

13. Rajput VD, Minkina TM, Behal A, Sushkova SN, Mandzhieva S, Singh R, Gorovtsov A, Tsitsuashvili VS, Purvis WO, Ghazaryan KA, Movsesyan HS. Effects of zinc-oxide nanoparticles on soil, plants, animals and soil organisms: A review. Environmental Nanotechnology, Monitoring & Management. 2018 May 1;9:76-84. https://doi.org/10.1016/j.enmm.2017.12.006

14. Gohari G, Mohammadi A, Akbari A, Panahirad S, Dadpour MR, Fotopoulos V, Kimura S. Titanium dioxide nanoparticles (TiO2 NPs) promote growth and ameliorate salinity stress effects on essential oil profile and biochemical attributes of Dracocephalum moldavica. Scientific reports. 2020 Jan 22;10(1):912. https://doi.org/10.1038/s41598-020-57794-1

15. Fatima, F., Hashim, A. & Anees, S. Efficacy of nanoparticles as nanofertilizer production: a review. *Environ Sci Pollut Res* 28, 1292–1303 (2021). https://doi.org/10.1007/s11356-020-11218-9

16. Nazim M, Li X, Anjum S, Ahmad F, Ali M, Muhammad M, Shahzad K, Lin L, Zulfiar U. Silicon nanoparticles: a novel approach in plant physiology to combat drought stress in arid environments. Biocatalysis and Agricultural Biotechnology. 2024 Apr 25:103190. https://doi.org/10.1016/j.bcab.2024.103190

17. Soliemanzadeh A, Fekri M. Effects of green iron nanoparticles on iron changes and phytoavailability in a calcareous soil. Pedosphere. 2021 Oct 1;31(5):761-70. https://doi.org/10.1016/S1002-0160(21)60035-8

18. Yuan P, Tan D, Annabi-Bergaya F. Properties and applications of halloysite nanotubes: recent research advances and future prospects. Applied Clay Science. 2015 Aug 1;112:75-93. https://doi.org/10.1016/j.clay.2015.05.001

19. Wu X, Liu C, Qi H, Zhang X, Dai J, Zhang Q, Zhang L, Wu Y, Peng X. Synthesis and adsorption properties of halloysite/carbon nanocomposites and halloysite-derived carbon nanotubes. Applied Clay Science. 2016 Jan 1;119:284-93. https://doi.org/10.1016/j.clay.2015.10.029

20. Fakhruddin K, Hassan R, Khan MU, Allisha SN, Abd Razak SI, Zreaqat MH, Latip HF, Jamaludin MN, Hassan A. Halloysite nanotubes and halloysite-based composites for biomedical applications. Arabian Journal of Chemistry. 2021 Sep 1;14(9):103294. https://doi.org/10.1016/j.arabjc.2021.103294

21. Luo P, Zhao Y, Zhang B, Liu J, Yang Y, Liu J. Study on the adsorption of Neutral Red from aqueous solution onto halloysite nanotubes. Water research. 2010 Mar 1;44(5):1489-97. https://doi.org/10.1016/j.watres.2009.10.042

22. Rozhina E, Batasheva S, Miftakhova R, Yan X, Vikulina A, Volodkin D, Fakhrullin R. Comparative cytotoxicity of kaolinite, halloysite, multiwalled carbon nanotubes and graphene oxide. Applied Clay Science. 2021 May 1;205:106041. https://doi.org/10.1016/j.clay.2021.106041

23. Chen L, Guo Z, Lao B, Li C, Zhu J, Yu R, Liu M. Phytotoxicity of halloysite nanotubes using wheat as a model: seed germination and growth. Environmental Science: Nano. 2021;8(10):3015-27. https://doi.org/10.1039/D1EN00507C

24. Francis AP, Devasena T. Toxicity of carbon nanotubes: A review. Toxicology and industrial health. 2018 Mar;34(3):200-10. https://doi.org/10.1177/0748233717747472

25. Masoud AR, Alakija F, Perves Bappy MJ, Mills PA, Mills DK. Metallizing the surface of halloysite nanotubes—a review. Coatings. 2023 Mar 2;13(3):542. https://doi.org/10.3390/coatings13030542

26. Biddeci G, Spinelli G, Colomba P, Di Blasi F. Nanomaterials: a review about halloysite nanotubes, properties, and application in the biological field. International Journal of Molecular Sciences. 2022 Sep 29;23(19):11518. https://doi.org/10.3390/ijms231911518

27. Metal A. Metal Oxide Nanoparticles in Bone Tissue Repair Ghazal Shineh, Mohammadmahdi Mobaraki, Elham Afzali, Femi Alakija, Zeinab Jabbari Velisdeh, David K Mills. Biomedical Materials & Devices. 2024;2(5).

28. Shineh G, Mobaraki M, Afzali E, Alakija F, Velisdeh ZJ, Mills DK. Antimicrobial metal and metal oxide nanoparticles in bone tissue repair. Biomedical Materials & Devices. 2024 Sep;2(2):918-41. https://doi.org/10.1007/s44174-024-00159-3

29. Masoud AR, Velisdeh ZJ, Bappy MJ, Pandey G, Saberian E, Mills DK. Cellulose-Based Nanofibers in Wound Dressing. Biomimetics. 2025 May 23;10(6):344. https://doi.org/10.3390/biomimetics10060344

30. Şahin CB, İşler N. The impact of foliar applied zinc and iron on quality of soybean. Journal of Plant Nutrition. 2023 Aug 9;46(13):2977-89. https://doi.org/10.1080/01904167.2022.2161389

31. Yruela I. Copper in plants. Brazilian journal of plant physiology. 2005;17:145-56. https://doi.org/10.1590/S1677-04202005000100012

32. Zahidah KA, Kakooei S, Ismail MC, Raja PB. Halloysite nanotubes as nanocontainer for smart coating application: A review. Progress in Organic Coatings. 2017 Oct 1;111:175-85. https://doi.org/10.1016/j.porgcoat.2017.05.018

33. Huber DM, Jones JB. The role of magnesium in plant disease. Plant and soil. 2013 Jul;368:73-85. https://doi.org/10.1007/s11104-012-1476-0

34. Senbayram M, Gransee A, Wahle V, Thiel H. Role of magnesium fertilisers in agriculture: plant–soil continuum. Crop and Pasture Science. 2015 Dec 21;66(12):1219-29. https://doi.org/10.1071/CP15104

35. Mordike BL, Ebert TJ. Magnesium: properties—applications—potential. Materials Science and Engineering: A. 2001 Apr 15;302(1):37-45. https://doi.org/10.1016/S0921-5093(00)01351-4

36. Oka R, Uwai K, Toyoguchi T, Shiraishi T, Nakagawa Y, Takeshita M. Stability of magnesium oxide products in single-dose packages. Iryo Yakugaku (Japanese Journal of Pharmaceutical Health Care and Sciences). 2007;33(12):1013-9. https://doi.org/10.5649/jjphcs.33.1013

37. Suliman AA, El-Dewiny CY, Soliman MK, Salem SS. Investigation of the Effects of Applying Bio-Magnesium Oxide Nanoparticle Fertilizer to Moringa Oleifera Plants on the Chemical and Vegetative Properties of the Plants' leaves. Biotechnology Journal. 2025 Mar;20(3):e202400536. https://doi.org/10.1002/biot.202400536

38. Kanjana D. Foliar application of magnesium oxide nanoparticles on nutrient element concentrations, growth, physiological, and yield parameters of cotton. Journal of Plant Nutrition. 2020 Jul 31;43(20):3035-49. https://doi.org/10.1080/01904167.2020.1799001

39. Machin J, Navas A. Soil pH changes induced by contamination by magnesium oxides dust. Land Degradation & Development. 2000 Jan;11(1):37-50. https://doi.org/10.1002/(SICI)1099-145X(200001/02)11:1%3C37::AID-LDR366%3E3.0.CO;2-8

40. Ibrahim MM, Lin H, Chang Z, Li Z, Riaz A, Hou E. Magnesium-doped biochars increase soil phosphorus availability by regulating phosphorus retention, microbial solubilization and mineralization. Biochar. 2024 Jul 18;6(1):68. https://doi.org/10.1007/s42773-024-00360-z

41. Samaraweera H, Palansooriya KN, Dissanayaka PD, Khan AH, Sillanpää M, Mlsna T. Sustainable phosphate removal using Mg/Ca-biochar hybrids: Current trends and future outlooks. https://doi.org/10.26434/chemrxiv-2023-vm51x

42. Wilcox GE, Pfeiffer CL. Temperature effect on seed germination, seedling root development and growth of several vegetables. Journal of plant nutrition. 1990 Nov 1;13(11):1393-403. https://doi.org/10.1080/01904169009364161

43. Ceylan Y, Kutman UB, Mengutay M, Cakmak I. Magnesium applications to growth medium and foliage affect the starch distribution, increase the grain size and improve the seed germination in wheat. Plant and soil. 2016 Sep;406:145-56. https://doi.org/10.1007/s11104-016-2871-8

44. Vishal B, Kumar PP. Regulation of seed germination and abiotic stresses by gibberellins and abscisic acid. Frontiers in plant science. 2018 Jun 20;9:838. https://doi.org/10.3389/fpls.2018.00838

45. Pan M, Yau PC, Lee KC, Man HY. Effects of different fertilizers on the germination of tomato and cucumber seeds. Water, Air, & Soil Pollution. 2022 Jan;233(1):25. https://doi.org/10.1007/s11270-021-05494-5

46. Szőllősi R, Molnár Á, Kondak S, Kolbert Z. Dual effect of nanomaterials on germination and seedling growth: Stimulation vs. phytotoxicity. Plants. 2020 Dec 10;9(12):1745. https://doi.org/10.3390/plants9121745

47. Kleijnen JP. Response surface methodology. InHandbook of simulation optimization 2014 Sep 18 (pp. 81-104). New York, NY: Springer New York. https://doi.org/10.1007/978-1-4939-1384-8_4

48. Ferreira SC, Bruns RE, Ferreira HS, Matos GD, David JM, Brandão GC, da Silva EP, Portugal LA, Dos Reis PS, Souza AS, Dos Santos WN. Box-Behnken design: An alternative for the optimization of analytical methods. Analytica chimica acta. 2007 Aug 10;597(2):179-86. https://doi.org/10.1016/j.aca.2007.07.011

49. Peixoto AF, Fernandes AC, Pereira C, Pires J, Freire C. Physicochemical characterization of organosilylated halloysite clay nanotubes. Microporous and Mesoporous Materials. 2016 Jan 1;219:145-54. https://doi.org/10.1016/j.micromeso.2015.08.002

50. Lvov Y, Wang W, Zhang L, Fakhrullin R. Halloysite clay nanotubes for loading and sustained release of functional compounds. Advanced Materials. 2016 Feb;28(6):1227-50. https://doi.org/10.1002/adma.201502341

51. Mao Y, Chai X, Zhong M, Zhang L, Zhao P, Kang Y, Guo J, Yang X. Effects of nitrogen and magnesium nutrient on the plant growth, quality, photosynthetic characteristics, antioxidant metabolism, and endogenous hormone of Chinese kale (Brassica albograbra Bailey). Scientia Horticulturae. 2022 Sep 20;303:111243. https://doi.org/10.1016/j.scienta.2022.111243

52. Dağhan HA. Effects of TiO2 nanoparticles on maize (Zea mays L.) growth, chlorophyll content and nutrient uptake. Applied ecology and environmental research. 2018;16. 10.15666/aeer/1605_68736883

53. Lvov YM, Shchukin DG, Mohwald H, Price RR. Halloysite clay nanotubes for controlled release of protective agents. ACS nano. 2008 May 27;2(5):814-20. https://doi.org/10.1021/nn800259q

54. Dewi IA, Sarfat MS. Nanoclay Composites: Innovative Agrochemical Delivery Systems for Enhanced Crop Production. https://doi.org/10.1039/9781837677313-00065

55. Hong J, Wang C, Wagner DC, Gardea-Torresdey JL, He F, Rico CM. Foliar application of nanoparticles: mechanisms of absorption, transfer, and multiple impacts. Environmental Science: Nano. 2021;8(5):1196-210. https://doi.org/10.1039/D0EN01129K

56. Zhao L, Lu L, Wang A, Zhang H, Huang M, Wu H, Xing B, Wang Z, Ji R. Nano-biotechnology in agriculture: use of nanomaterials to promote plant growth and stress tolerance. Journal of agricultural and food chemistry. 2020 Jan 31;68(7):1935-47. https://doi.org/10.1021/acs.jafc.9b06615

57. Velisdeh ZJ, Najafpour Darzi G, Poureini F, Mohammadi M, Sedighi A, Bappy MJ, Ebrahimifar M, Mills DK. Turning Waste into Wealth: Optimization of Microwave/Ultrasound-Assisted Extraction for Maximum Recovery of Quercetin and Total Flavonoids from Red Onion (Allium cepa L.) Skin Waste. Applied Sciences. 2024 Oct 11;14(20):9225. https://doi.org/10.3390/app14209225

58. Velisdeh ZJ, Najafpour GD, Mohammadi M, Poureini F. Optimization of sequential microwave-ultrasound-assisted extraction for maximum recovery of quercetin and total flavonoids from red onion (Allium cepa L.) skin wastes. arXiv preprint arXiv:2104.06109. 2021 Apr 13. https://doi.org/10.48550/arXiv.2104.06109

59. Gent MP. Dynamic carbohydrate supply and demand model of vegetative growth: response to temperature, light, carbon dioxide, and day length. Agronomy. 2018 Feb 16;8(2):21. https://doi.org/10.3390/agronomy8020021

60. Singh S, Husen A. Role of nanomaterials in the mitigation of abiotic stress in plants. InNanomaterials and plant potential 2019 Mar 2 (pp. 441-471). Cham: Springer International Publishing. https://doi.org/10.1007/978-3-030-05569-1_18

61. Allahverdiyeva Y, Aro EM. Photosynthetic responses of plants to excess light: mechanisms and conditions for photoinhibition, excess energy dissipation and repair. InPhotosynthesis: plastid biology, energy conversion and carbon assimilation 2011 Aug 3 (pp. 275-297). Dordrecht: Springer Netherlands. https://doi.org/10.1007/978-94-007-1579-0_13

62. Camejo D, Rodríguez P, Morales MA, Dell'Amico JM, Torrecillas A, Alarcón JJ. High temperature effects on photosynthetic activity of two tomato cultivars with different heat susceptibility. Journal of plant physiology. 2005 Mar 14;162(3):281-9. https://doi.org/10.1016/j.jplph.2004.07.014

63. Joussein E, Petit S, Churchman J, Theng B, Righi D, Delvaux BJ. Halloysite clay minerals—a review. Clay minerals. 2005 Dec;40(4):383-426.

64. Wamelink GW, Frissel JY, Krijnen WH, Verwoert MR, Goedhart PW. Can plants grow on Mars and the moon: a growth experiment on Mars and moon soil simulants. PLoS One. 2014 Aug 27;9(8):e103138. https://doi.org/10.1371/journal.pone.0103138

65. Duri LG, Caporale AG, Rouphael Y, Vingiani S, Palladino M, De Pascale S, Adamo P. The potential for lunar and martian regolith simulants to sustain plant growth: a multidisciplinary overview. Frontiers in Astronomy and Space Sciences. 2022 Jan 4;8:747821. https://doi.org/10.3389/fspas.2021.747821